\documentclass[12pt]{article}

\usepackage{amsmath}
\usepackage{amssymb}

\def\rdots{\mathinner{\mkern1mu\raise1pt\vbox{\kern1pt\hbox{.}}\mkern2mu
   \raise4pt\hbox{.}\mkern2mu\raise7pt\hbox{.}\mkern1mu}}
\newcommand{\Z}{{\rm Z\kern-.35em Z}}
\newcommand{\bP}{{\rm I\kern-.15em P}}
\newcommand{\Q}{\kern.3em\rule{.07em}{.65em}\kern-.3em{\rm Q}}
\newcommand{\R}{{\rm I\kern-.15em R}}
\newcommand{\h}{{\rm I\kern-.15em H}}
\newcommand{\C}{\kern.3em\rule{.07em}{.65em}\kern-.3em{\rm C}}
\newcommand{\T}{{\rm T\kern-.35em T}}

\newcommand{\be}{\begin{equation}}
\newcommand{\ee}{\end{equation}}

\newcommand{\la}{\lambda}

\newcommand{\ra}{\rightarrow}

\newcommand{\al}{\alpha}

\begin{document}

\openup 1.5\jot
\centerline{Dimer $\la_d$ Expansion, A Contour Integral}
\centerline{Stationary Point Argument}
\bigskip

\bigskip

\vspace{1in}
\centerline{Paul Federbush}
\centerline{Department of Mathematics}
\centerline{University of Michigan}
\centerline{Ann Arbor, MI 48109-1043}
\centerline{(pfed@umich.edu)}

\vspace{1in}

\centerline{\underline{Abstract}}

In the development of a presumed asymptotic expansion for $\la_d$ in previous work, a basic step involved extracting the asymptotic behavior of a sum as being dominated by the largest term in the sum.  But this argument only is valid when the terms in the sum are positive.  In the case terms in the sum are not all positive (some of the $\bar J_n$ are negative) we replace the `largest term' argument by the route of this paper.  The sum is replaced by a contour integral.  We then find a stationary point of the integrand that we conjecture dominates the asymptotic behavior of the integral.  

\vfill\eject

An asymptotic expansion for $\la_d$ in powers of $1/d$ has been studied in a series of papers, [1] - [4].  In [1] we extracted the $N \ra \infty$ asymptotic behavior of the sum in equation (33) therein
\be   Z^* = \sum_{\al_i} \beta \big(N, \sum\; i\; \al_i \; \big) \bar J^{\al_1}_1 \cdots \bar J^{\al_{s+1}}_{s+1} \cdot 
\frac{N^{\sum \; \al_i }}{\al_1 ! \cdots \al_{s+1} !}  \ee
(with very slight change in notation) by seeking the largest term in this sum.  This argument depended on the terms in the sum being positive.  For a while we feared the expansion of $\la_d$ only applied where only positive $\bar J_n$ were involved.  But, as promised after equation (35) of [1], an alternate argument will be given herein yielding the same expressions for the $\la_d$ asymptotic series.

We replace (1) by the asymptotically equivalent formula:
\be Z^* = \left( \frac 1{2\pi i} \right)^{s+1} \oint dz_1 \cdots \oint dz_{s+1} \; \frac 1{z_1 \cdots z_{s+1}} \bigg[ \sum^{\mu N}_{\al_1 = -\mu N} \cdot \cdot \sum^{\mu N}_{\al_{s+1} = -\mu N} \beta(N,\sum\;i\;\al_i) \ \cdot \ee
\[  \cdot\  z^{\al_1}_1 \cdots z^{\al_{s+1}}_{s+1} \bigg] e^{N\; \sum^{s+1}_{i=1} \; \bar J_i \; \frac 1{z_i}}  \]
In both expressions we are seeking contributions from $\bar J_1, ..., \bar J_{s+1}$, and $\mu$ and the $\bar J_i$ should satisfy
\be	|\bar J_i| \ << \mu << 1 \ . \ee

We make a formal change of variables (in the $N \ra \infty$ limit)
\begin{eqnarray}
\al_i &\longrightarrow& N \rho_i \\
\Delta \; \al_i &\longrightarrow& N d\rho_i \nonumber
\end{eqnarray}
to arrive at 
  \be Z^* \cong \left( \frac 1{2\pi i} \right)^{s+1} \oint dz_1 \cdots \oint dz_{s+1} \; \frac 1{z_1 \cdots z_{s+1}} \bigg[ \int^\mu_{-\mu} Nd\rho_1 \cdot \cdot \int^\mu_{-\mu} Nd\rho_{s+1}  \beta(N,\sum_i Ni\rho_i) \ \cdot \ee
\[	\cdot \ e^{N\; \sum^{s+1}_{i=1} \rho_i ln(z_i)}\bigg] e^{N\; \sum^{s+1}_{i=1} \; \bar J_i \; \frac 1{z_i}}  \]
and substituting from the unnumbered equation between (33) and (34) in [1], (5) becomes
\be Z^* \cong \left( \frac 1{2\pi i} \right)^{s+1} \oint dz_1 \cdots \oint dz_{s+1} \; \frac 1{z_1 \cdots z_{s+1}}  \int^\mu_{-\mu} Nd\rho_1 \cdot \cdot \int^\mu_{-\mu} Nd\rho_{s+1} \ \cdot \ee
\[ \cdot \ \ e^{N\left\{\left[\left( \frac{1-2\sum_i \; \rho_i}{2} \right) ln (1-2\sum_i \; \rho_i) + \sum_i i\rho_i \right] + \sum_i \rho_i ln(z_i) +  \sum_i \bar J_i \; \frac1{z_i} \right\} } \ .\]

We conjecture that the $N \ra \infty$ behavior of (1), or (2), may be obtained from the behavior of the integrand of (6) at its stationary point.  We expect there will soon be a proof of this, perhaps it follows from a known result.  Applying the conjecture and differentiating the quantity in braces, $\{ \; \cdot \; \}$, with respect to $\rho_i$ and $z_i$ and setting all derivatives equal zero one gets exactly equations (34) - (36) of [1].

Our `contour integral stationary point' technique is easily seen to pass the following two tests.

\bigskip
\bigskip

\noindent
\underline{Test 1}  If all the $\bar J_i$ are non-negative then the result obtained agrees with the result of the `largest term' computation.

\noindent
\underline{Test 2}  If the function $\beta$ is set equal 1, this method yields the correct result
\be		Z^* \ = \ e^{N\sum \bar J_i} .  \ee

Proving the conjecture above may seem peremptory, but what one really needs is the result in the limit $s\ra \infty$, for the bounds actually holding for $\bar J_i$ (for large $d$).  We do not yet know what bounds hold for the $\bar J_i$, a seemingly very difficult task to find such.  Much work remains.

\bigskip
\bigskip
\bigskip

\noindent
\underline{ACKNOWLEDGMENT}  I hope this work on $\la_d$ is interesting enough that I get invited to give a seminar at Princeton, where I received much of my education.

\vfill\eject

\centerline{References}
\begin{itemize}
\item[[1]] Paul Federbush, ``Hidden Structure in Tilings, Conjectured Asymptotic Expansion for $\la_d$ in Multidimensional Dimer Problem", \ arXiv : 0711.1092V9 [math-ph].
\item[[2]] Paul Federbush, ``Dimer $\la_d$ Expansion Computer Computations", arXiv: 0804.4220V1 [math-ph].
\item[[3]] Paul Federbush, ``Dimer $\la_3 = .453\pm.001$ and Some Other Very Intelligent Guesses", arXiv:0805.1195V1 [math-ph].
\item[[4]] Paul Federbush, ``Dimer $\la_d$ Expansion, Dimension Dependence of $\bar J_n$ Kernels",  arXiv:0806.1941V1 [math-ph].

\end{itemize}

\end{document}